\documentclass[aps,prl,preprint,nopacs,superscriptaddress]{revtex4}
\usepackage{graphicx}
\usepackage{verbatim}
\usepackage{relsize}
\usepackage{mathrsfs}
\usepackage{color}
\usepackage{colortbl}
\usepackage{ulem}
\usepackage{blindtext}
\usepackage{amsmath,amsfonts,amssymb}
\usepackage{graphicx}

\def \beq {\begin{equation}}
\def \eeq {\end{equation}}
\begin{document}

\title{Optical detection and manipulation of spontaneous gyrotropic electronic order in a transition-metal dichalcogenide semimetal\\}

\author{Su-Yang Xu$^*$}\affiliation {Department of Physics, Massachusetts Institute of Technology, Cambridge, Massachusetts 02139, USA}
\author{Qiong Ma$^*$}\affiliation{Department of Physics, Massachusetts Institute of Technology, Cambridge, Massachusetts 02139, USA}

\author{Yang Gao\footnote{These authors contributed equally to this work.}}\affiliation{Department of Physics, Carnegie Mellon University, Pittsburgh, Pennsylvania 15213, USA}
\author{Anshul Kogar}\affiliation {Department of Physics, Massachusetts Institute of Technology, Cambridge, Massachusetts 02139, USA}
\author{Alfred Zong}\affiliation {Department of Physics, Massachusetts Institute of Technology, Cambridge, Massachusetts 02139, USA}
\author{Andr\'es M. Mier Valdivia}\affiliation{Department of Physics, Massachusetts Institute of Technology, Cambridge, Massachusetts 02139, USA}
\author{Thao H. Dinh}\affiliation{Department of Physics, Massachusetts Institute of Technology, Cambridge, Massachusetts 02139, USA}
\author{Shin-Ming Huang}\affiliation{Department of Physics, National Sun Yat-sen University, Kaohsiung 80424, Taiwan}

\author{Bahadur Singh}\affiliation{SZU-NUS Collaborative Center and International Collaborative Laboratory of 2D Materials for Optoelectronic Science \& Technology, Engineering Technology Research Center for 2D Materials Information Functional Devices and Systems of Guangdong Province, College of Optoelectronic Engineering, Shenzhen University, ShenZhen 518060, China}\affiliation{Department of Physics, Northeastern University, Boston, Massachusetts 02115, USA}

\author{Chuang-Han Hsu}
\affiliation{Department of Electrical and Computer Engineering, National University of Singapore, Singapore 117683}

\author{Tay-Rong Chang}\affiliation{Department of Physics, National Cheng Kung University, Tainan 701, Taiwan}

\author{Jacob P.C. Ruff}\affiliation{CHESS, Cornell University, Ithaca, NY 14853, USA}

\author{Kenji Watanabe}

\affiliation{National Institute for Materials Science, Namiki 1 -1, Tsukuba, Ibaraki 305 -0044, Japan}
\author{Takashi Taniguchi}

\affiliation{National Institute for Materials Science, Namiki 1 -1, Tsukuba, Ibaraki 305 -0044, Japan}

\author{Hsin Lin}
\affiliation{Institute of Physics, Academia Sinica, Taipei 11529, Taiwan}

\author{Goran Karapetrov}\affiliation{Department of Physics and Department of Materials Science and Engineering, Drexel University, 3141 Chestnut St., Philadelphia, PA 19104}

\author{Di Xiao}\affiliation{Department of Physics, Carnegie Mellon University, Pittsburgh, Pennsylvania 15213, USA}

\author{Pablo Jarillo-Herrero$^{\dag}$}\affiliation {Department of Physics, Massachusetts Institute of Technology, Cambridge, Massachusetts 02139, USA}

\author{Nuh Gedik\footnote{Corresponding authors (emails): pjarillo@mit.edu and gedik@mit.edu}}\affiliation {Department of Physics, Massachusetts Institute of Technology, Cambridge, Massachusetts 02139, USA}

%\date{\today}
\maketitle

\textbf{The observation of chirality is ubiquitous in nature. Contrary to intuition, the population of opposite chiralities is surprisingly asymmetric at fundamental levels \cite{wagniere2008chirality, feringa1999absolute}. Examples range from parity violation in the subatomic weak force \cite{wagniere2008chirality} to the homochirality in essential biomolecules \cite{feringa1999absolute}. The ability to achieve chirality-selective synthesis (chiral induction) is of great importance in stereochemistry, molecular biology and pharmacology \cite{feringa1999absolute}. In condensed matter physics, a crystalline electronic system is geometrically chiral when it lacks any mirror plane, space inversion center or roto-inversion axis \cite{flack2003chiral}. Typically, the geometrical chirality is predefined by a material's chiral lattice structure, which is fixed upon the formation of the crystal. By contrast, a particularly unconventional scenario is the gyrotropic order \cite{van2011chirality, hosur2013kerr, orenstein2013berry, pershoguba2013proposed, fu2015parity}, where chirality spontaneously emerges across a phase transition as the electron system breaks the relevant symmetries of an originally achiral lattice. Such a gyrotropic order, proposed as the quantum analogue of the cholesteric liquid crystals, has attracted significant interest \cite{van2011chirality, hosur2013kerr, orenstein2013berry, pershoguba2013proposed, fu2015parity, ishioka2010chiral, iavarone2012evolution, castellan2013chiral, xia2008polar, hildebrand2018local, varma2014gyrotropic, zenker2013chiral, ganesh2014theoretical, gradhand2015optical}. However, to date, a clear observation and manipulation of the gyrotropic order remain challenging. We report the realization of optical chiral induction and the observation of a gyrotropically ordered phase in the transition-metal dichalcogenide semimetal $1T$-TiSe$_2$. We show that shining mid-infrared circularly polarized light near the critical temperature leads to the preferential formation of one chiral domain. As a result, we are able to observe an out-of-plane circular photogalvanic current, whose direction depends on the optical induction. Our study provides compelling evidence for the spontaneous emergence of chirality in the correlated semimetal TiSe$_2$ \cite{di1976electronic}. Such chiral induction provides a new way of optical control over novel orders in quantum materials.}

In the presence of strong correlations, the behavior of electrons in a metal can signficantly deviate from a weakly-interacting Fermi gas, forming a wide range of complex, broken symmetry phases \cite{fradkin2010nematic1}. Recent theoretical works have highlighted their analogy with classical liquids \cite{fradkin2010nematic1}, where a rich set of liquid crystalline phases that exhibit varying degrees of symmetry breaking and transport properties have been observed. A well-studied case is the nematic order \cite{fradkin2010nematic1, ronning2017electronic, harter2017parity}, i.e., the spontaneous emergence of rotational anisotropy. In classical liquids, this is known as the nematic liquid crystal, whereas in quantum materials, it has recently been observed in quantum Hall systems, ruthenates, high temperature superconductors \cite{fradkin2010nematic1}, heavy-fermion superconductors \cite{ronning2017electronic}, and correlated metallic pyrochlores \cite{harter2017parity}. Another fascinating case is the gyrotropic order, i.e., the spontaneous emergence of geometrical chirality. In classical liquids, this is known as the cholesteric liquid crystal. In quantum materials, despite fundamental interest \cite{van2011chirality, hosur2013kerr, orenstein2013berry, pershoguba2013proposed, fu2015parity, ishioka2010chiral, iavarone2012evolution, castellan2013chiral, xia2008polar, hildebrand2018local, varma2014gyrotropic, zenker2013chiral, ganesh2014theoretical}, a clear observation of the gyrotropic order remains lacking.

For a condensed matter system, one route to obtain chirality is to look for low-symmetry materials that naturally crystalize in lattice-structures that are free of any mirror, inversion and roto-inversion symmetries, i.e., chiral crystals \cite{flack2003chiral}. Representative chiral crystal materials include quartz, tellurium, carbon nanotubes and twisted bilayer graphene. On the other hand, a highly unconventional route is the gyrotropic order: An electronic system can spontaneously break space-inversion, mirror reflection, and roto-inversion symmetries of the originally achiral lattice, and, as a result, gain a definitive chirality. Such a gyrotropic order \cite{van2011chirality, hosur2013kerr, orenstein2013berry, pershoguba2013proposed, fu2015parity} was suggested in $1T$-TiSe$_2$ \cite {ishioka2010chiral, iavarone2012evolution, castellan2013chiral} and cuprates \cite{xia2008polar, he2011single} based on scanning tunneling microscopy (STM) and polar Kerr rotation measurements. However, later studies \cite{armitage2014constraints, hosur2015kerr, hildebrand2018local} including a recent STM experiment \cite{hildebrand2018local} on $1T$-TiSe$_2$ argued that the existing experimental evidence remains inconclusive. Moreover, the manipulation of the emergent chirality has not been achieved. On the other hand, theoretical studies have found that the gyrotropic order can arise from a number of electronic instabilities \cite{Note}, including chiral charge-density waves \cite{van2011chirality, hosur2013kerr, orenstein2013berry, wang2014polar}, loop current order \cite{varma2014gyrotropic, pershoguba2013proposed}, odd parity electronic (Pomeranchuk) instability \cite{fu2015parity, pomeranchuk1958stability}, and electron-phonon driven instability \cite{zenker2013chiral}. Moreover, since the gyrotropic order must be odd under parity, it may serve as a strong precursor for odd parity Cooper pairing \cite{kozii2015odd, wang2016topological}, offering a novel mechanism for unconventional superconductivity \cite{ganesh2014theoretical, hosur2013kerr, orenstein2013berry, varma2014gyrotropic, kozii2015odd}. In addition, recent theoretical studies \cite{chang2018universal, de2017quantized} predicted that geometrical chirality can guarantee the existence of topological Weyl fermions. Thus the manipulation of emergent chirality in a gyrotropic phase may allow for the realization of highly tunable Berry curvature and topological properties. These new physics further call for the unambiguous demonstration and manipulation of the gyrotropic order in quantum materials.

 One challenge is to identify a clear experimental signature for geometrical chirality in metals. For insulating chiral crystals or molecules, chirality can be measured by optical activity \cite{orenstein2013berry, gradhand2015optical}, which manifests as the rotation of the linear polarization plane as light travels through a nonmagnetic chiral medium. However, this is difficult in metallic bulk materials as they do not transmit light in bulk samples. Another challenge is to align chiral domains, because otherwise the signal of the chirality probe will be cancelled by the small domains of opposite chirality. The ferroelectric, ferromagnetic, and nematic domains can be aligned by electric, magnetic, or uniaxial-strain fields, respectively, because these fields couple asymmetrically to opposite domains. However, none of them can couple asymmetrically to opposite chiralities. In the present work, we have carefully addressed these challenges. First, we identified a particular type of photocurrent as a clear signature of chirality in metallic electron systems. The circular photogalvanic effect (CPGE) generates a photocurrent whose direction depends on the chirality of light \cite{olbrich2009observation, mciver2012control}. Here, we show that the out-of-plane CPGE current under normal incident light, i.e., the CPGE current parallel to the light propagation direction, serves as a clear signature of the emergence of geometrical chirality in a metallic electron system \cite{de2017quantized}. Second, we realized a chirality-selective stimulus. Among all the external stimuli available, circularly polarized light is a rare example that has a definitive chirality \cite{feringa1999absolute}. We demonstrate the chiral induction for the gyrotropic order by mid-infrared circularly polarized light.

\begin{figure*}[t]
\includegraphics[width=17.5cm]{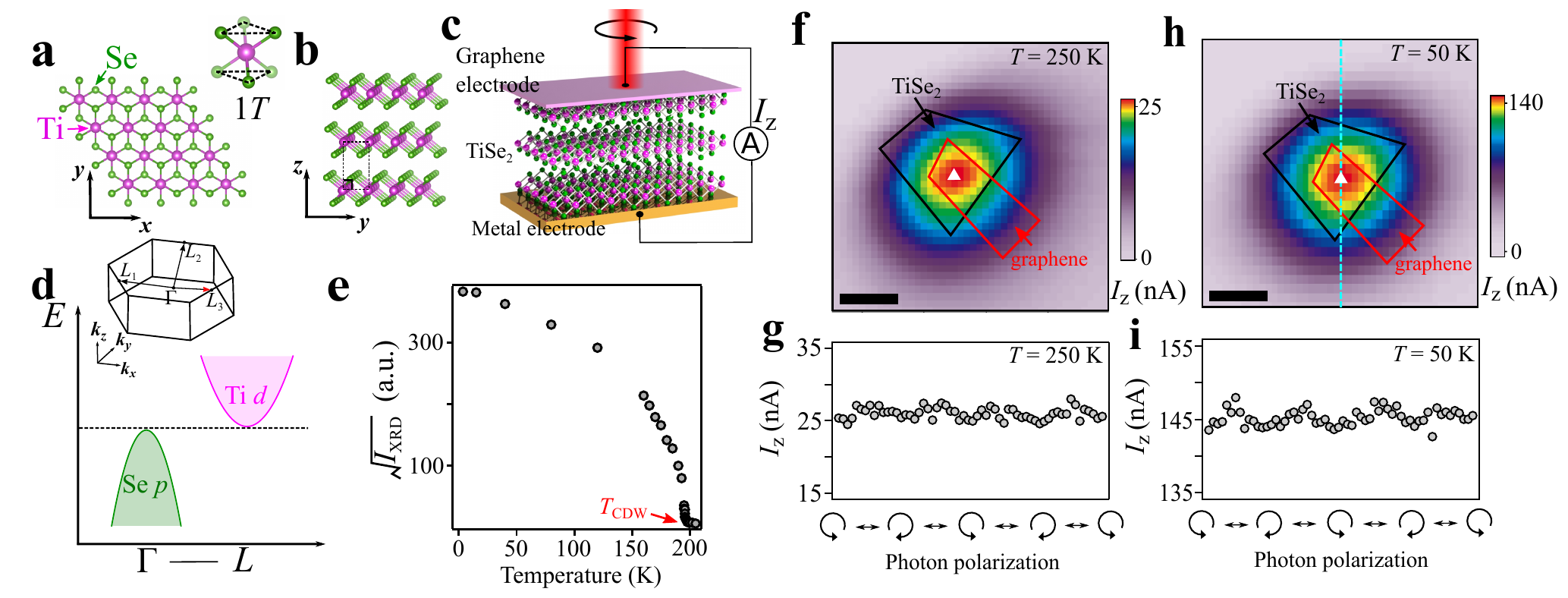}
\caption{{\bf Crystal structure and basic characterizations of $1T$-TiSe$_2$.} \textbf{a,b,} Crystal structure of the $1T$-TiSe$_2$. The space group is $P$-$3m1$ (centrosymmetric). \textbf{c,} Our vertical photoactive devices consist of a TiSe$_2$ flake sandwiched by a transparent graphene electrode on the top and a metal electrode on the bottom. Normal incident mid-infrared light is applied on the device and the out-of-plane photocurrent $I_{z}$ is collected. \textbf{d,} The normal state of $1T$-TiSe$_2$ features a semimetal band structure with a vanishing, indirect band gap, where the valence band maximum and the conduction band minimum are located at the $\Gamma$ and $L$ points of the bulk Brillouin zone. Inset: There are three inequivalent $L$ points ($L_1$, $L_2$ and $L_3$). In momentum space, the $2\times2\times2$ CDW state corresponds to three independent CDWs whose propagation vectors connect $\Gamma$ to $L_1$, $L_2$ and $L_3$, respectively. \textbf{e,} Square root of the x-ray diffraction intensity of superlattice peak (Miller index $(\frac{1}{2}, \frac{1}{2}, \frac{23}{2})$) due to the $2\times2\times2$ CDW. The critical temperature of the CDW $T_{\textrm{CDW}}$ is determined to be around $198$ K. \textbf{f,} The out-of-plane photocurrent $I_{z}$ with linear polarization measured as a function of the beam spot location in real space (below we refer to this kind of measurements as a photocurrent spatial map) at $T=250$ K. Scare bar in panels (\textbf{f}, \textbf{h}): $25$ $\mu$m. \textbf{g,} The $I_{z}$ measured as a function of light polarization with the beam spot parked near the device center (the white triangles in panel (\textbf{f})) at $T=250$ K. \textbf{h,i,} The same as panels (\textbf{f,g}) but at $T=50$ K.}
\label{Fig1}
\end{figure*}

The layered  transition-metal dichalcoginide $1T$-TiSe$_2$ (Figs.~\ref{Fig1}\textbf{a,b}) is a correlated semimetal. It has attracted great interest because of the rich variety of novel electronic properties \cite{kidd2002electron, rossnagel2002charge, Morosan2006, cercellier2007evidence, li2007semimetal, rohwer2011collapse, porer2014non, chen2015charge, li2016controlling, kogar2017signatures}. It undergoes a commensurate charge-density wave (CDW) transition at $T_{\textrm{CDW}}\sim200$ K, leading to a simple ($2\times2\times2$) superlattice \cite{di1976electronic, holt2001x} (Figs.~\ref{Fig1}\textbf{a,b}). Its vanishing, indirect band gap (Fig.~\ref{Fig1}\textbf{d}) \cite{kidd2002electron, rossnagel2002charge, cercellier2007evidence, li2007semimetal, rohwer2011collapse, porer2014non}, low carrier concentration and strong Coulomb interactions create optimal conditions for excitonic condensation \cite{halperin1968possible,cercellier2007evidence, rohwer2011collapse, kogar2017signatures}. Moreover, superconductivity and incommensurate CDW emerge by chemical doping \cite{Morosan2006} in the bulk and by electrostatic gating in ultra-thin flakes \cite{li2016controlling}. Furthermore, a truly 2D CDW has been observed recently in monolayer TiSe$_2$ \cite{chen2015charge}.

Our experimental setup involves a mid-infrared scanning photocurrent microscope equipped with a continuous-wave CO$_2$ laser ($\lambda=10.6$ $\mu$m, $\hbar\omega=117$ meV), which allows us to measure the mid-infrared photocurrent as a function of beam spot location, light polarization and temperature (Fig.~\ref{Fig1}\textbf{c}). In order to probe the out-of-plane photocurrents $I_{z}$, we have fabricated vertical photoactive devices consisting of a TiSe$_2$ flake sandwiched by a transparent graphene electrode on the top and a metal electrode on the bottom (Fig.~\ref{Fig1}\textbf{c}). The thickness of the TiSe$_2$ flakes is comparable to the laser skin depth ($\simeq200$ nm) calculated from the optical conductivity at $\lambda=10.6$ $\mu\textrm{m}$ \cite{li2007semimetal}. This ensures that a significant vertical fraction of the flake is photoactive. The critical temperature of the $2\times2\times2$ CDW phase transition of our TiSe$_2$ samples was determined by X-ray diffraction to be around $198$ K (Fig.~\ref{Fig1}\textbf{e}). We have measured four devices from two batches of crystals and obtained consistent results.

We characterize the basic photoresponse of our devices. Figure~\ref{Fig1}\textbf{f} presents the photocurrent spatial map at $T=250$ K, i.e., $I_{z}$ as a function of beam spot location in real space. Figure~\ref{Fig1}\textbf{g} shows polarization dependence of $I_{z}$ with the beam spot parked at the sample center. These data reveal a polarization-independent $I_{z}$ signal that roughly spans the entire area of the TiSe$_2$ sample. We identify this signal as the commonly observed photo-thermoelectric current \cite{mciver2012control}. Laser-heating induces a temperature gradient, which in turn leads to a thermoelectric current across a heterojunction with different Seebeck coefficients, such as the top graphene/TiSe$_2$ interface. Lowering the temperature ($T=50$ K in Figs.~\ref{Fig1}\textbf{h,i}) gives qualitatively similar results, except that the magnitude of this polarization-independent photo-thermoelectric signal grows larger. For the datasets presented above, the sample was cooled down without being illuminated by laser. 

\begin{figure*}[t]
\includegraphics[width=17cm]{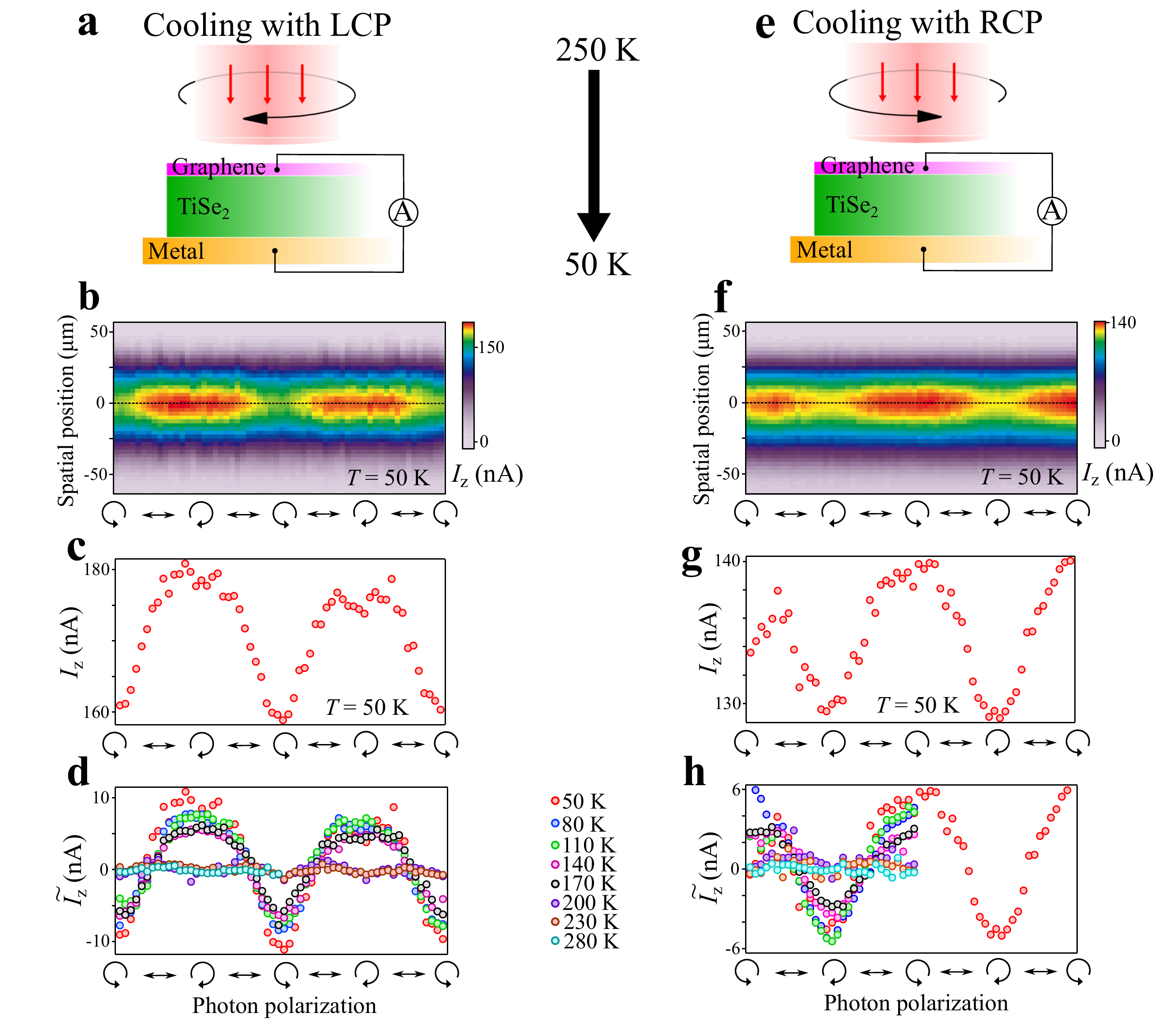}
\caption{{\bf Observation of an emergent out-of-plane circular photogalvanic current upon chiral induction.} \textbf{a,} We shine left circularly polarized (LCP) light on the sample while lowering its temperature from $T=250$ K to $50$ K. \textbf{b,} The out-of-plane photocurrent $I_z$ at $T=50$ K as a function of the light polarization (horizontal axis) and the beam spot location (vertical axis). The trajectory of the beam spot location is denoted by the blue dashed line in Fig.~\ref{Fig1}\textbf{h}. \textbf{c,} $I_z$ at $T=50$ K as a function of polarization. A clear out-of-plane CPGE is observed. \textbf{d,} $\tilde{I}_z$ at different temperatures as we warm the sample up from $50$ K toward room temperature. Here $\tilde{I}_z$ means the polarization-dependent part (the CPGE) of ${I}_z$. The polarization-independent part of ${I}_z$ for this data set is presented in Extended data Fig.~\ref{EF1}. We observe that the CPGE persists up to $170$ K and then vanishes rather abruptly. \textbf{e-h,} Same as panels (\textbf{a-d}), except that we shine right circularly polarized (LCP) light on the sample while lowering its temperature from $T=250$ K to $50$ K. The emergent CPGE is found to be reversed. }
\label{Fig2}
\end{figure*}

We now shine circularly polarized light on the sample during the entire cooling process. Surprisingly, a markedly different situation is observed. As shown in Fig.~\ref{Fig2}\textbf{a}, we shine left circularly polarized (LCP) light on the sample while lowering its temperature from $T=250$ K. Upon reaching low temperature ($T=50$ K), we measure $I_{z}$ as a function of polarization with the beam spot parked at the sample center (Fig.~\ref{Fig2}\textbf{c}). We observe a clear polarization dependence: $I_{z}$ is maximum for LCP and minimum for right circular polarization (RCP). This pattern clearly demonstrates the emergence of an out-of-plane CPGE photocurrent. Apart from the CPGE, the polarization-independent photo-thermoelectric signal remains present. We further vary the beam spot location along the blue dashed line in Fig.~\ref{Fig1}\textbf{h} and study $I_{z}$ as a function of polarization at each location along this line. As shown in Fig.~\ref{Fig2}\textbf{b}, clear CPGE is observed as long as the beam spot is on the sample. By increasing temperature step by step, we observe in Fig.~\ref{Fig2}\textbf{d} that the CPGE persists up to $170$ K and then vanishes rather abruptly. On the other hand, as expected, the polarization-independent photo-thermoelectric signal evolves smoothly with increasing temperature and remains finite all the way up to $280$ K (Extended data Fig.~\ref{EF1}). The sharply different behaviors, including the temperature dependence and the reliance on circularly polarized illumination during cooling, further distinguish the CPGE from the photo-thermoelectric signal. We repeat the measurements described above but change the light polarization during cooling to RCP (Figs.~\ref{Fig2}\textbf{e-h}). Remarkably, the sign of the out-of-plane CPGE is reversed (Figs.~\ref{Fig2}\textbf{f-h}). Other behaviors such as the temperature dependence (Fig.~\ref{Fig2}\textbf{h}) and the spatial distribution (Fig.~\ref{Fig2}\textbf{f}) are qualitatively similar.

\begin{figure*}[t]
\includegraphics[width=12cm]{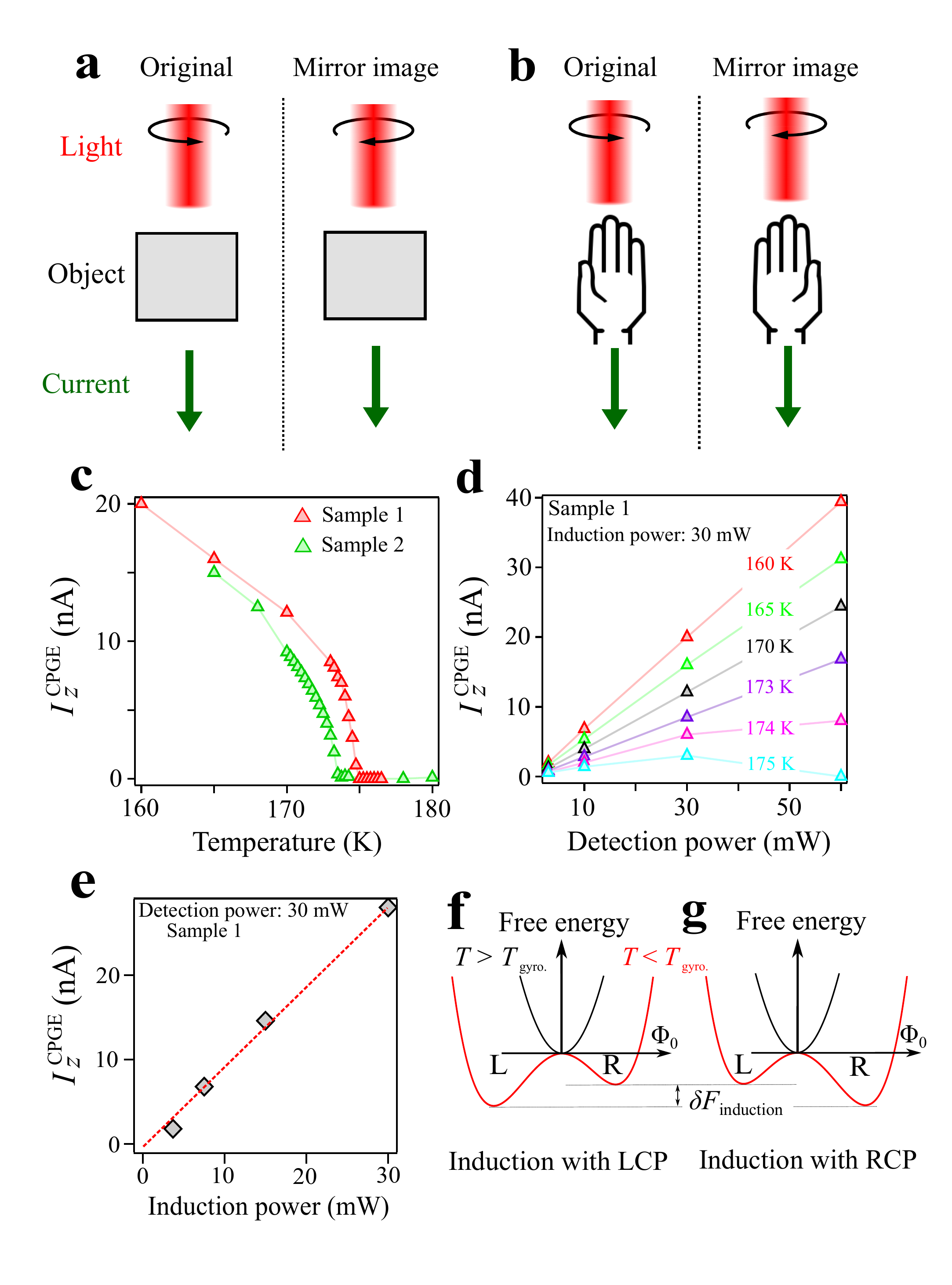}
\caption{{\bf Systematic temperature and laser power dependences of the chiral induction and the spontaneous emergent gyrotropic order.} \textbf{a,b,} We present an intuitive physical picture to establish the connection between the out-of-plane CPGE and geometrical chirality. LCP and RLP refer to left and right circular polarization. The current refer to the photocurrent induced by shining circularly polarized light. See the main text for relevant descriptions. \textbf{c,} After cooling down to $T=150$ K with LCP light, we measure the $I_{z}^{\textrm{CPGE}}$ with small temperature increments. The results (Fig.~\ref{Fig3}\textbf{c}) clearly uncover the onset critical temperature ($T_{\textrm{gyro.}}\simeq174$ K) for the emergent geometrical chirality. Sample 1 is the sample presented in Figs.~\ref{Fig1} and ~\ref{Fig2}. Sample's systematic photocurrent data are presented in Extended Data Fig.~\ref{EF2}.}
\label{Fig3}
\end{figure*}

We now establish the connection between the out-of-plane CPGE and geometrical chirality. In the main text, we adopt an intuitive physical picture while detailing the mathematical proof in SI.II. Starting from a mirror symmetric, achiral system, we assume that shining LCP light leads to an out-of-plane photocurrent $I_{z}$ parallel to the light propagation direction (Fig.~\ref{Fig3}\textbf{a}). We then perform a mirror transformation to the entire conceptual experiment (Fig.~\ref{Fig3}\textbf{a}). Under mirror reflection, LCP changes to RCP, the mirror symmetric, achiral crystal remains invariant, and the $I_{z}$ (a vector parallel to the mirror plane) also remains invariant. Therefore, $I_{z}$ under LCP and RCP are identical for a mirror symmetric, achiral crystal (Fig.~\ref{Fig3}\textbf{a}). In other words, the out-of-plane CPGE ($I_{z}^{\textrm{CPGE}}=I_{z}^{\textrm{RCP}}-I_{z}^{\textrm{LCP}}$) is forbidden by the mirror symmetry of the crystal. The key to the above symmetry constraint is that the mirror transformation flips the chirality of light but keeps the crystal invariant. Such a symmetry constraint is lifted in a chiral system. As light and the crystal are both chiral, a mirror transformation flips their chirality simultaneously. Therefore, mirror symmetry only forces the $I_{z}$ under LCP in a left-handed crystal and $I_{z}$ under RCP in a reft-handed crystal to be identical. (Fig.~\ref{Fig3}\textbf{b}). On the other hand, for a crystal with a fixed chirality (e.g. a left-handed chiral crystal), there is no symmetry that can relate the $I_{z}^{\textrm{RCP}}$ and $I_{z}^{\textrm{LCP}}$.  Therefore, the out-of-plane CPGE is symmetry allowed in a chiral crystal but forbidden in an achiral crystal. Conversely, we conclude that the observation of the out-of-plane CPGE in Fig.~\ref{Fig2} provides a clear experimental signature for the emergent geometrical chirality in the low temperature state of $1T$-TiSe$_2$. 

We further systematically study the out-of-plane CPGE $I_{z}^{\textrm{CPGE}}$ as a function of temperature and laser power. In this paragraph, we focus on presenting the data. How these data reveal the physics of the gyrotropic phase transition and chiral induction will be discussed below. We study how $I_{z}^{\textrm{CPGE}}$ vanishes with increasing temperature: After cooling down to $T=150$ K with LCP light, we measure the $I_{z}^{\textrm{CPGE}}$ with small temperature increments. The results (Fig.~\ref{Fig3}\textbf{c}) uncover the onset critical temperature ($T_{\textrm{gyro.}}$) for the emergence of geometrical chirality. For all samples tested, we found $T_{\textrm{gyro.}}\simeq174$ K. We then study $I_{z}^{\textrm{CPGE}}$'s laser power dependence. In Fig.~\ref{Fig3}\textbf{d}, we fix the induction-power (the laser power used for chiral induction), and measure the $I_{z}^{\textrm{CPGE}}$ as a function of the detection-power (the laser power used to measure the CPGE) at various temperatures. At temperatures significantly lower than $T_{\textrm{gyro.}}$, $I_{z}^{\textrm{CPGE}}$ depends linearly on the detection-power. On the other hand, at temperatures close to $T_{\textrm{gyro.}}$, $I_{z}^{\textrm{CPGE}}$ shows a non-monotonic pattern. With increasing detection-power, $I_{z}^{\textrm{CPGE}}$ first increases then decreases.  In Fig.~\ref{Fig3}\textbf{e}, we fix the detection-power, and measure the $I_{z}^{\textrm{CPGE}}$ as a function of the induction-power at $T=100$ K. The results show that $I_{z}^{\textrm{CPGE}}$ grows approximately linearly with induction-power.
 In Extended Data Fig.~\ref{EF2}, we present measurements on another type of device that can detect both the in-plane and out-of-plane photocurrents independently, which confirms the observation of the out-of-plane CPGE and the chiral induction. 

Finally, we discuss how our data uncover physics of the gyrotropic phase transition and chiral induction in $1T$-TiSe$_2$. Our thorough exploration shows that chiral induction can be achieved by shining circularly polarized light while cooling across $T_{\textrm{gyro.}}$. On the other hand, directly applying circularly polarized light at low temperatures is unable to reverse the emergent chirality  (for laser powers within our experimental capability). This suggests that the chiral domains are extremely stiff at low temperatures \cite{liang2017orthogonal}. Indeed, since the CPGE itself is measured by varying light polarization between left and right, the detection of the CPGE at low temperatures requires the chiral domains to be ``frozen''. This is further supported by the observed detection-power dependence in Fig.~\ref{Fig3}\textbf{d}: The linear detection-power profile at low temperatures agrees with the CPGE's second order optoelectronic nature. On the other hand, the non-monotonic profile near $T_{\textrm{gyro.}}$ may be a sign that the chiral domains are partially reversed by the detection light. However, we note that, based on our current data, this cannot be distinguished from the other possibility that the gyrotropic order is melted by laser heating. 

We analyze our observations within the framework of the Landau phase transition theory \cite{toledano1987landau}, paying particular attention to the broken symmetries. Generically, the order parameter for a gyrotropic phase transition can be expressed as $\Phi_{\textrm{gyro.}}=\Phi_0(T)f(\mathbf{r})$, where $f(\mathbf{r})$ is odd under inversion, mirror and roto-inversion symmetries, and $\Phi_0(T)$ is a real number whose value depends on temperature. As temperature is lowered across $T_{\textrm{gyro.}}$, $\Phi_0$ becomes nonzero, which, as a result, breaks the inversion, mirror and roto-inversion symmetries of the system. $\Phi_0<0$ and $\Phi_0>0$ correspond to left and right chiral domains, which are energetically degenerate in the absence of optical chiral induction. We now aim to identify a proper form for the coupling between circularly polarized light and the chiral domains (denoted as $\delta F_{\textrm{induction}}$). The circularly polarized light is characterized by its time-dependent, spatially rotating electric field. Since $\delta F_{\textrm{induction}}$ must be time-independent, it cannot be linearly proportional to $\mathbf{E}$. Therefore, we have performed systematic symmetry analyses based on our experimental observations, and we identified $\delta F_{\textrm{induction}}$, to the lowest order of the electric field of light, in the form of $\delta F_{\textrm{induction}}=\Phi_{\textrm{gyro.}}\big[(\mathbf{E}\times \frac{\partial\mathbf{E}}{ \partial z})\cdot \hat{z}\big]$, where $\mathbf{E}$ is the electric field of the light ($\mathbf{E}=|\mathbf{E}|\big[\cos(\frac{2\pi}{\lambda}z-\omega t)\hat{x}\pm \sin(\frac{2\pi}{\lambda}z-\omega t) \hat{y}$\big]). One can check that $\big[(\mathbf{E}\times \frac{\partial\mathbf{E}}{ \partial z})\cdot \hat{z}\big]$, where $\mathbf{E}$ is time-independent and it flips sign as one reverses light chirality. Moreover, $\big[(\mathbf{E}\times \frac{\partial\mathbf{E}}{ \partial z})\cdot \hat{z}\big]$ is also odd under any mirror reflection, space inversion and roto-inversion. Therefore, $\delta F_{\textrm{induction}}=\Phi_{\textrm{gyro.}}\big[(\mathbf{E}\times \frac{\partial\mathbf{E}}{ \partial z})\cdot \hat{z}\big]$ as a whole is totally symmetric and transforms as a scalar. This is crucial because otherwise $\delta F_{\textrm{induction}}$ cannot enter the free energy \cite{toledano1987landau}. One can check that this form of $\delta F_{\textrm{induction}}$ tilts the total free energy landscape in opposite ways depending on light chirality (Figs.~\ref{Fig3}\textbf{f,g}). Furthermore, this coupling term $\delta F_{\textrm{induction}}$ suggests a linear dependence on the optical induction power ($\delta F_{\textrm{induction}}\propto E^2$), which is well supported by our data in Fig.~\ref{Fig3}\textbf{e}.

We further analyze the gyrotropic order.  One crucial aspect is its relationship with the $2\times2\times2$ CDW phase transition. It is known that, across $T_{\textrm{CDW}}$, translational symmetry is broken. The question is whether additional point group symmetries (e.g. inversion, mirror and roto-inversion) are also broken at $T_{\textrm{CDW}}$. Our systematic theoretical analyses show that, irrespective of the underlying mechanism (Peierls or exciton condensation), across $T_{\textrm{CDW}}$, inversion symmetry must remain intact. In other words, we are able to theoretically prove that the $2\times2\times2$ CDW phase transition at $T_{\textrm{CDW}}$ cannot turn $1T$-TiSe$_2$ to a chiral state at once. This theoretical finding is consistent with our experimental observation that the critical temperature for the gyrotropic order is lower than that of the CDW transition ($T_{\textrm{gyro.}}<T_{\textrm{CDW}}$). Therefore, our data and analyses demonstrate that $1T$-TiSe$_2$ goes first from the high temperature normal phase to a $2\times2\times2$ achiral CDW phase at $T_{\textrm{CDW}}$ and then to a chiral phase at $T_{\textrm{gyro.}}$. Below $T_{\textrm{gyro.}}$, the gyrotropic order and the CDW order coexist. The microscopic origin of the gyrotropic order requires further investigations. Other important open questions include the mechanism for the chiral induction as well as the chiral domain wall dynamics. These questions point to promising future directions both in theory and in experiments.

Nonetheless, our measurements and analyses clearly demonstrate the gyrotropic order and chiral induction, paving the way for exciting experimental possibilities \cite{Morosan2006, li2016controlling, kozii2015odd, wang2016topological}. Our work also establishes the mid-infrared CPGE photocurrent as a powerful probe for novel broken symmetry states in metals. In particular, the mid-infrared photon energy corresponds to important energy scales such as the CDW gap in TiSe$_2$ and the pseudo-gap in cuprates. Furthermore, the CPGE's second order effect nature is uniquely sensitive to odd-parity order parameters \cite{fu2015parity, harter2017parity}. Therefore, mid-infrared CPGE may provide new insights to the pseudo-gap state of cuprates and the newly observed insulating and superconducting order in ``magic angle'' twisted bilayer graphene \cite{cao2018unconventional}. Finally, the chiral induction observed here demonstrates a new way of optical control over novel orders in quantum materials. It may be used to control other unusual electronic states with geometrical chirality such as skyrmions and helical magnetic orders in spin-orbit coupled quantum magnets.

\bibliographystyle{naturemag}
\bibliography{Topological_and_2D_12062017}

\begin{thebibliography}{10}
\bibitem{Note}
\bibinfo{title}{{We note that the gyrotropic phase transition is fundamentally
  defined by the corresponding symmetry breaking. Specifically, across a
  gyrotropic phase transition, a system changes from achiral to chiral by
  breaking inversion, mirror and roto-inversion symmetries. Such a phase
  transition can be further characterized by its underlying mechanism. For
  instance, a chiral charge-density wave is one specific type of gyrotropic
  phase transition, where the chirality arises from certain chiral modulation
  of the charge-density waves. See SI. VI.2 for a detailed discussion of the
  possible microscopic mechanisms}} .
\end{thebibliography}

\vspace{0.5cm}
\textbf{Methods}

\textbf{Sample fabrication:} First, bottom metal (PdAu) contacts were deposited on standard Si/SiO$_2$ substrates. These substrates with bottom contacts were then transferred into an argon environment glovebox with the levels of water and oxygen levels below 0.1 ppm. TiSe$_2$ was exfoliated inside the glovebox, and flakes with a thickness about a few hundreds nanometers were identified.
The selected TiSe$_2$ flake was first picked up by a poly(biphenol A carbonate) (PC) stamp and directly transferred onto the substrate. The PC was then dissolved. Subsequently, h-BN, graphene and h-BN flakes were picked up one by one, and the whole stack was then transferred on top of the TiSe$_2$ flake. The graphene served as transparent top electrode and it touches the side electrode for wire bonding. The lower h-BN layer worked as the edge spacer between the top electrode graphene and the bottom metal. The top h-BN covered the active area (the overlapping area between top graphene, TiSe$_2$ and bottom contact) to prevent the device from degradation. 

\textbf{X-ray diffraction} X-ray measurements were performed at the CHESS-A2 synchrotron beamline, using $38.5$ keV photons.  Incident photon energy was selected using a perfect HPHT diamond $<111>$ double-crystal monochromator, with a narrow incident bandwidth $\frac{dE}{E} < 1e^{-4}$.  The incident beam on sample was ~150 microns tall and ~600 microns wide.  The crystal was mounted on a temperature-controlled copper post inside a closed-cycle cryostat, with a temperature stability better than $\pm 0.1$ K and absolute temperature calibration better than 1 K. CDW peak intensities were collected using a $4$-circle diffractometer and a high-dynamic-range photon-counting area detector.

\textbf{Mid-infrared scanning photocurrent microscopy:} The laser source is a temperature-stablized CO$_2$ laser ($\lambda = 10.6$ $\mu$m or $\hbar\omega =  117$ meV). A focused beam spot (beam waist $\simeq 25$ $\mu$m) is scanned (using a two axis piezo-controlled scanning mirror) over the entire sample and the current is recorded at the same time to form a two-dimensional map of photocurrent as a function of spatial positions. Reflected light from the sample is collected to form a simultaneous reflection image of the sample. The absolute location of the photo-induced signal is therefore found by comparing the photocurrent map to the reflection image. The light polarization is modulated by rotating a quarter-wave plate. The CPGE is an intrinsic photocurrent response \cite{sipe2000, hosur2011, morimoto2016}. As a second order effect, the CPGE current depends quadratically on the electric field of light (here we refer to the light used to measure the CPGE). The in-plane CPGE under oblique or normal incidence has been studied in a wide range of materials 
\cite{ ivchenko2008, Planck2018, olbrich2009observation, mciver2012control,yuan2014, dhara2015, ma2017, de2017quantized, lim2018, ji2018, xu2018}. Any intrinsic CPGE requires the material to be inversion breaking \cite{sipe2000, hosur2011, morimoto2016,  ivchenko2008, Planck2018, olbrich2009observation, mciver2012control,yuan2014, dhara2015, ma2017, de2017quantized, lim2018, ji2018, xu2018}.

\vspace{0.5cm}

\textbf{Data availability:} The data that support the plots within this paper and other findings of this study are available from the corresponding author upon reasonable request.

\textbf{Acknowledgement:} We thank Xiaodong Xu, Valla Fatemi, Edbert J. Sie and Shiang Fang for important discussions. NG, SYX, AG and GAZ acknowledge support from DOE, BES DMSE (data taking and analysis), and the Gordon and Betty Moore Foundations EPiQS Initiative through Grant GBMF4540 (manuscript writing). Work in the PJH  group was supported through AFOSR grant FA9550-16-1-0382 (fabrication and measurement), as well as through the Center for the Advancement of Topological Semimetals, an Energy Frontier Research Center funded by the U.S. Department of Energy Office of Science, Office of Basic Energy Sciences (data analysis), and the Gordon and Betty Moore Foundation’s EPiQS Initiative through Grant GBMF4541 to PJH. This work made use of the Materials Research Science and Engineering Center Shared Experimental Facilities supported by the National Science Foundation (NSF) (Grant No. DMR-0819762). The work at Drexel University was supported by NSF under Grant No. ECCS-1711015. Research conducted at the Cornell High Energy Synchrotron Source (CHESS) is supported by the NSF \& NIH/NIGMS via NSF award DMR-1332208. Work at CMU was supported by the Department of Energy, Basic Energy Sciences, Materials Sciences and Engineering Division, Grant No. DE-SC0012509. KW and TT acknowledge support from the Elemental Strategy Initiative conducted by the MEXT, Japan, JSPS KAKENHI Grant Numbers JP18K19136 and the CREST (JPMJCR15F3), JST. T.R.C was supported by the Ministry of Science and Technology under MOST Young Scholar Fellowship: MOST Grant for the Columbus Program NO. 107-2636-M-006-004-, National Cheng Kung University, Taiwan, and National Center for Theoretical Sciences (NCTS), Taiwan. S.M.H. acknowledges support by the Ministry of Science and Technology (MoST) in Taiwan under Grant No. 105-2112-M-110-014-MY3. H.L. acknowledges Academia Sinica, Taiwan for the support under Innovative Materials and Analysis Technology Exploration (AS-iMATE-107-11). 
 
\textbf{Author contributions:} NG and PJH supervised the project. SYX and QM conceived the experiment. SYX and QM performed photocurrent measurements and analysed the data. AK, JPCR, SYX and QM performed X-ray diffraction measurements. AMMV, TD, QM and SYX fabricated the devices. YG and SYX performed theoretical analysis under the supervision of DX. CHH, SMH, BS, TRC performed DFT calculations under the supervision of HL. GK grew TiSe$_2$ crystals. KW and TT grew the bulk hBN single crystals. AZ made significant contributions to the symmetry analysis and the overall presentation. SYX, QM, PJH and NG wrote the manuscript with input from all authors. 

\textbf{Competing financial interests:} The authors declare no competing financial interests.

\clearpage

\clearpage

\addtocounter{figure}{-1}
\begin{figure*}[t!]
\caption{\textbf{d,} We fix the induction-power (the laser power used for chiral induction), and measure the $I_{z}^{\textrm{CPGE}}$ as a function of the detection-power (the laser power used to measure the CPGE) at various temperatures. At temperatures significantly lower than $T_{\textrm{gyro.}}$, $I_{z}^{\textrm{CPGE}}$ depends linearly on the detection-power. On the other hand, at temperatures close to $T_{\textrm{gyro.}}$, $I_{z}^{\textrm{CPGE}}$ shows a non-monotonic pattern. \textbf{e,} We fix the detection-power, and measure the $I_{z}^{\textrm{CPGE}}$ as a function of the induction-power at $T=100$ K. Squares are data points. The red dotted line is a linear fit. \textbf{f,} Schematic illustration of the free energy as a function of the order parameter $\Phi_0$ (see main text). Below $T_{\textrm{gyro.}}$, the free energy develops a double-well shape, where $\Phi_0<$ and $\Phi_0>0$ correspond to the left and right chiral domains, respectively. The coupling between the chiral domain and the circular polarized light, characterized by $\delta F_{\textrm{induction}}$, can tilt the double well profile, which therefore makes one chiral domain energetically more favorable than the other.}
\end{figure*}

\setcounter{figure}{0}
\renewcommand{\figurename}{\textbf{Extended Data Fig.}}

\clearpage
\begin{figure*}[t]
\includegraphics[width=14cm]{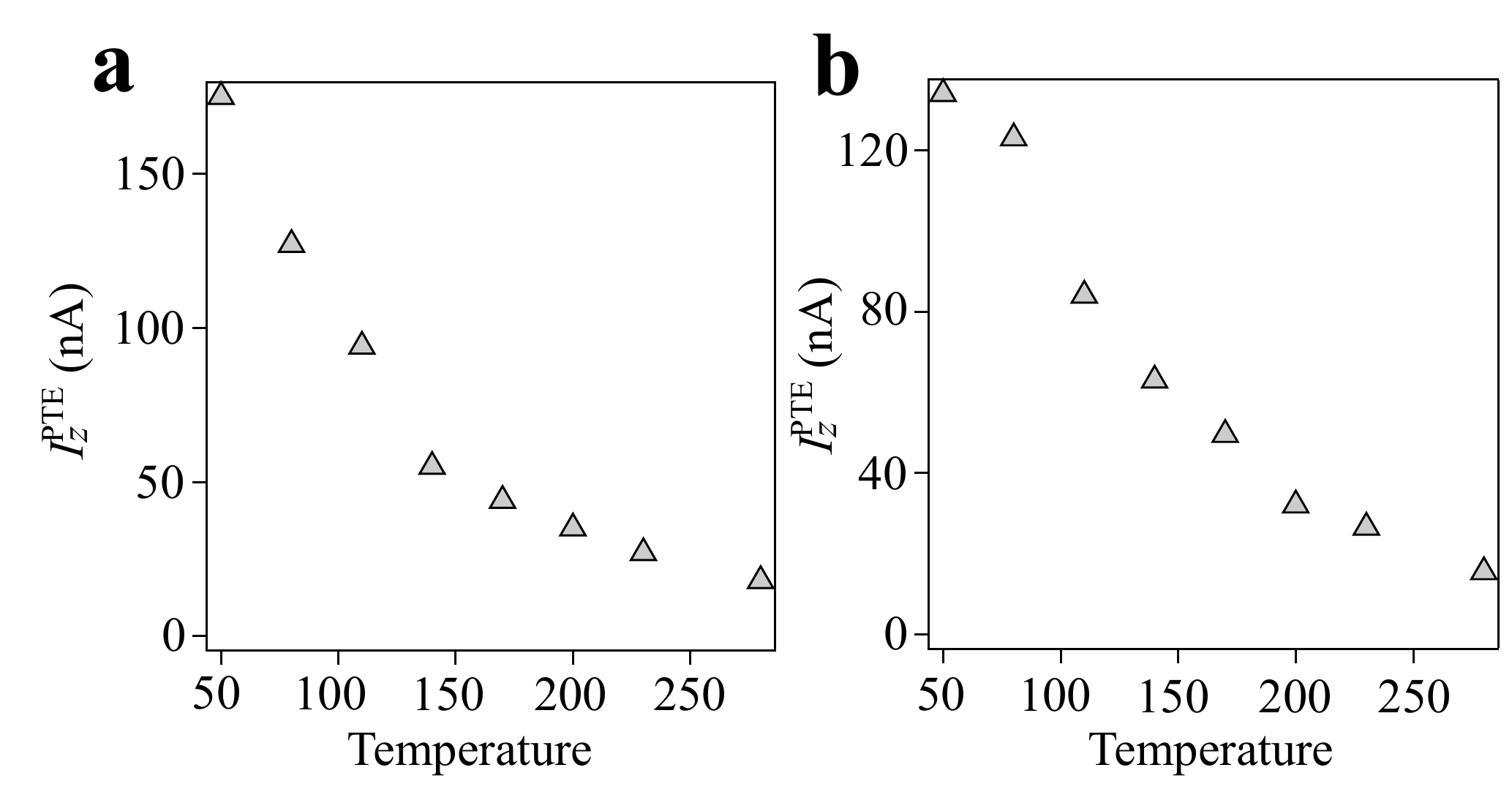}
\caption{Polarization-independent (photo-thermoelectric (PTE)) $I^{\textrm{PTE}}_z$ data at different temperatures as we warm the sample up from $50$ K toward room temperature. The data here come from the same photocurrent measurements as Figs.~\ref{Fig2}\textbf{d,h}. The data here are the polarization-independent component whereas Figs.~\ref{Fig2}\textbf{d,h} are the polarization-dependent component of the photocurrent data.}
\label{EF1}
\end{figure*}

\begin{figure*}[t]
\includegraphics[width=11cm]{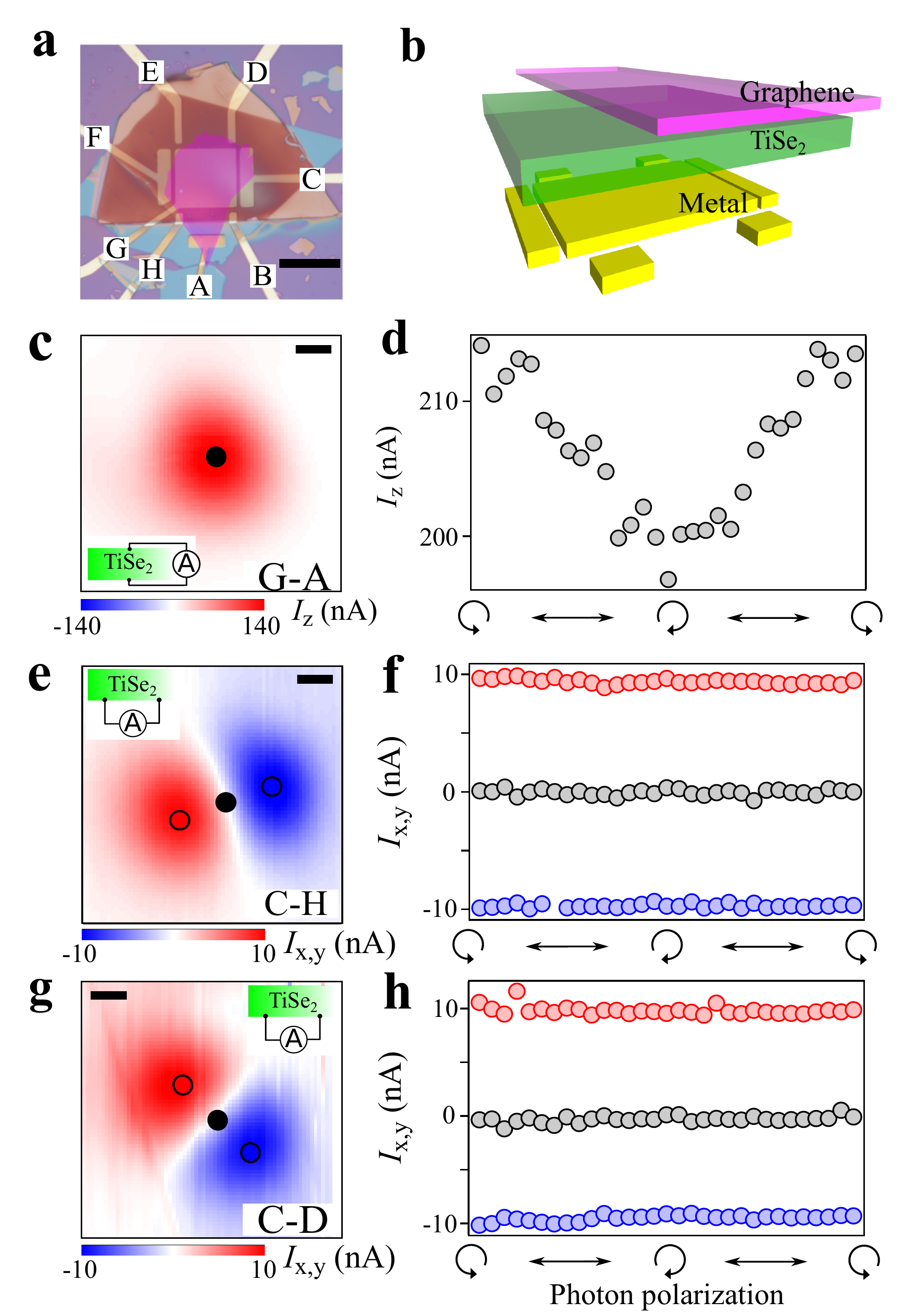}
\caption{{\bf Sharply contrasting behaviors for in-plane and out-of-plane photocurrents as a result of the gyrotropic order in $1T$-TiSe$_2$} \textbf{a,b,} Optical image and schematic illustration of a TiSe$_2$ photoactive device, which can independently detect both the in-plane and out-of-plane photocurrents. The multiple electrodes are enumerated by capital letters. \textbf{c,} The out-of-plane photocurrent $I_z$ spatial map at $T=100$ K measured between the electrodes G-A. \textbf{d,} The polarization-dependent $I_z$ data at $T=100$ K. The $I_{z}$ on this device is consistent with the results on previous devices (Figs. 2 and 3). }
\label{EF2}
\end{figure*}
\addtocounter{figure}{-1}
\begin{figure*}[t!]
\caption{The spatial map (panel (\textbf{c})) shows a single peak that roughly covers the device. This peak consists of two independent signals: The polarization-independent background arises from photo-thermoelectric effect and exists at all temperatures; The polarization-dependent CPGE signal (panel (\textbf{d})) relies on the chiral induction and only appears at low temperatures. \textbf{e,f,} and \textbf{g,h,} Same as panels \textbf{c,d} but for the in-plane photocurrent $I_{x,y}$ between the electrodes C-H and C-D, respectively. The $I_{x,y}$ shows distinctly different properties. No in-plane CPGE is observed (panels (\textbf{f,h})). The polarization-independent signal shows a bipolar spatial configuration (panels (\textbf{e,g})). Specifically, $I_{x,y}$ changes sign as the beam spot is scanned from one contact to the other. Such a bipolar spatial configuration, as widely observed in other in-plane photocurrent studies \cite{mciver2012control}, further confirms that the polarization-independent signals are photo-thermoelectric currents at the contact/sample junctions. Scar bars: 25 $\mu$m. All data in this figure are collected after RCP chiral induction with an induction power of $30$ mW.}
\end{figure*}

\end{document}